# A Lattice Study of Spin and Flavour Symmetry in Heavy Quark Physics

UKQCD collaboration, presented by Henning Hoeber [a] *

[a]Department of Physics & Astronomy, University of Edinburgh, Edinburgh EH9 3JZ, Scotland

We present a first test of heavy quark spin-flavour symmetries in matrix elements for semi-leptonic decays $\bar{B} \to D^* l\bar{\nu}$ and $\bar{B} \to Dl\bar{\nu}$. We show that $O(1/m_Q)$ corrections are small at masses around the charm for form factors protected by Luke's Theorem, but are of order 30-40 % for $h_V$ and $h_V/h_{A_1}$.

## 1. Introduction

In the limit of infinitely heavy $b$ and $c$ quarks an exact spin-flavour symmetry relates the six form factors describing the semi-leptonic decays $\bar{B} \to Dl\bar{\nu}$ and $\bar{B} \to D^*l\bar{\nu}$ to a single universal function, $\xi(v \cdot v')$, the Isgur-Wise (IW) function[1].

The form factors are defined as follows :

$$\frac{\langle D(v')|V^\mu|\bar{B}(v)\rangle}{\sqrt{m_j m_i}} = h_+(v \cdot v')(v+v')^\mu$$
$$+ h_-(v \cdot v')(v-v')^\mu$$

$$\frac{\langle D^*(v',\epsilon)|V^\mu|\bar{B}(v)\rangle}{\sqrt{m_j m_i}} = ih_V(v \cdot v')\,\epsilon^{\mu\nu\lambda\sigma}\,\epsilon^*_\nu v'_\lambda v_\sigma$$

$$\frac{\langle D^*(v',\epsilon)|A^\mu|\bar{B}(v)\rangle}{\sqrt{m_j m_i}} = (1+v\cdot v')h_{A_1}(v\cdot v')\epsilon^{*\mu}$$
$$-\epsilon^* \cdot v\{h_{A_2}(v\cdot v')v^\mu + h_{A_3}(v\cdot v')v'^\mu\}, \quad (1)$$

where $v$ and $v'$ are the four-velocities of the initial and final mesons.

To be able to have any confidence in a determination of the CKM-matrix element $V_{cb}$[9][10] we need to reliably determine the leading symmetry breaking coefficients in an expansion in $1/m_Q$. The theoretical framework to include these corrections is provided by Heavy Quark Effective Theory. To parametrize the matrix elements at order $1/m_Q$ one needs to introduce four additional functions and a mass parameter[3]. Together with the IW function these functions describe the interactions of the light degrees of freedom in the heavy mesons and they are therefore only calculable non-perturbatively. Typically, Heavy Quark Symmetry (HQS) makes predictions for the normalisation of decay rates or ratios thereof at zero recoil, but the dependence on $q^2$ is not predicted.

A lattice calculation of the meson weak decay form factors provides a systematic study of HQS from first principles. In this calculation we use the Wilson action with an $O(a)$ improvement[2]. A different approach can be taken by deriving an effective action where the heavy degrees of freedom have been integrated out[4].

Recently, the complete expressions for meson weak decay amplitudes have been given to order $1/m_Q$ and to next-to-leading order in QCD perturbation theory[5]. For the form factors of equation (1) we have

$$h_i(v \cdot v') = \{1 + \beta_i + O(\frac{\Lambda_{QCD}}{m_Q})\}\xi_{ren}(v \cdot v') \quad (2)$$

where $\beta_i(m_{b,c}, v \cdot v')$ are short distance corrections. As we wish to display $1/m_Q$ symmetry breaking effects we show plots of

$$h_i^{rad.corr} \equiv \frac{h_i(v \cdot v')}{1 + \beta_i(m_{b,c}, v \cdot v')} \quad (3)$$

As defined in (2), $\xi_{ren}(v \cdot v')$ is renormalisation-group invariant and normalised to one at zero recoil.

## 2. Simulation details

These preliminary results are obtained from 60 quenched, $\beta = 6.2$, $24^3 \times 48$ lattices, corresponding to a scale of $a^{-1} = 2.73(5)$GeV, as set by the string tension[6]. Quark propagators were generated using an $O(a)$-improved action at two "bottom" hopping parameters, $\kappa_Q = 0.129$ and $0.121$, corresponding to spin averaged meson masses of 1.4 and 1.9GeV[7]; four "charm" hopping parameters, $\kappa_{Q'} = 0.133, 0.129, 0.125, 0.121$, and three "strange" hopping parameters, $\kappa_l =$

*The author is greatful for financial support from the Dewar and Ritchie Fund.

0.14144, 0.14226, 0.14262. The vector(axial) currents are $O(a)$-improved versions of $\bar{b}\gamma_\mu(\gamma_5)c$. The method for calculating the matrix elements is well established [8]. The use of non-conserved local currents is taken into account by introducing rescaling factors $Z_V$ and $Z_A$. The determination of these constants is described in [9] and [10].

Errors are calculated using a bootstrap procedure on 100 samples and take into account correlations between different timeslices.

By injecting momenta up to $\sqrt{2}(\frac{2\pi}{La})$ at the current and momenta $(0,0,0)$ and $(1,0,0)$ for the B-meson we vary $v \cdot v'$ in the range 1 to 1.3.

## 3. Flavour symmetry - form factors at maximum recoil

Our study of flavour symmetry is motivated by *Luke's Theorem* which states that there are no terms of order $1/m_Q$ in the hadronic matrix elements at zero recoil[3]. In terms of form factors this means that only $h_+(1)$ and $h_{A_1}(1)$ are protected against first order corrections, i. e. : $h_{A_1}(1) = 1 + O(\frac{1}{m_Q^2})$. Corrections of order $\frac{1}{m_Q^2}$ have been estimated to be less than 3% [5] and we may use these results to determine $\rho^2$ to good accuracy from either of these two form factors.

To explore flavour symmetry we exploit the fact that knowledge of the IW function at maximum recoil, $q^2 = 0$, determines its $q^2$ dependence completely. As we have computed three-point correlators over a wide range of $b$ and $c$ masses we may vary $v \cdot v'$ by variation of the heavy meson masses[11]. Define $\hat{y}$ by

$$\hat{y} \equiv v \cdot v'(q^2 = 0) = \frac{1}{2}(\frac{m_B}{m_{D^*}} + \frac{m_{D^*}}{m_B}). \quad (4)$$

For every mass combination we fit $h_{A_1}(v.v')$ to Stech's relativistic-oscillator parametrization(BSW) and interpolate to $\hat{y}$. These values can then again be fitted to the BSW form which parametrizes the IW function in terms of the slope parameter $-\rho^2 = \xi'_\rho(1)$:

$$\xi_\rho(v \cdot v') = \frac{2}{1 + v \cdot v'} \exp\{-(2\rho^2 - 1)\frac{v \cdot v' - 1}{v \cdot v' + 1}\} (5)$$

We perform a two parameter fit to $s\xi_\rho(v \cdot v')$ where the parameter $s$ is introduced to absorb uncertainties in the overall normalisation.

In Figure 1 we show a fit to the data for the heaviest of the light quarks, $\kappa_l = 0.14144$, a $b$

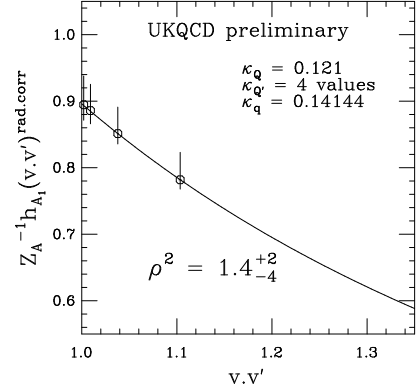

Figure 1. Isgur-Wise function, normalised to $Z_A^{-1}$. The slope is determined from a fit to data from different heavy quark masses at zero recoil.

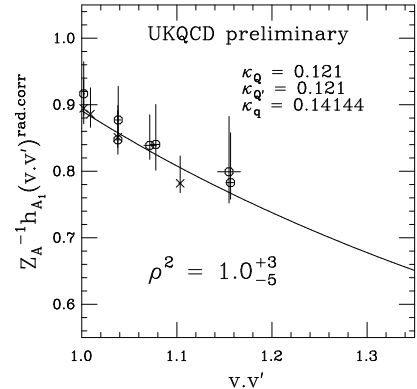

Figure 2. Isgur-Wise function, normalised to $Z_A^{-1}$, at fixed heavy quark mass(circles) and from the data from Fig. 1(crosses).

quark of $\kappa_Q = 0.121$ and all four values for the $c$ quark. Fitting these points to $s\xi_\rho(v \cdot v')$ we find $\rho^2 = 1.4 \pm {}^{2}_{4}$ whereas a fit to the data for the same light quark and degenerate heavy quarks of $\kappa_Q = 0.121$ yields $\rho^2 = 1.0 \pm {}^{3}_{5}$(Figure 2). Finding good agreement we conclude that flavour symmetry holds over the whole range of masses and that we are in the heavy quark limit at around the charm mass. Using the fact that $Z_A$ is essentially mass independent[10] this allows us to determine $\rho^2$ from a fit to the data for more than one mass combination. In Figure 3 we show a BSW fit to the data for the lighter of the two $b$ quarks and all four $c$ quark values. We find $\rho^2 = 1.2 \pm {}^{2}_{3}$, consistent with the above results.

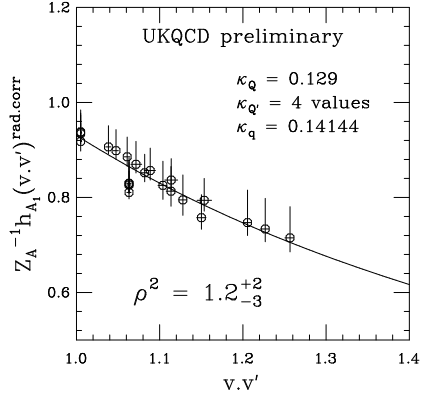

Figure 3. Isgur-Wise function, normalised to $Z_A^{-1}$. The slope is obtained from a BSW fit to data from different heavy quark masses.

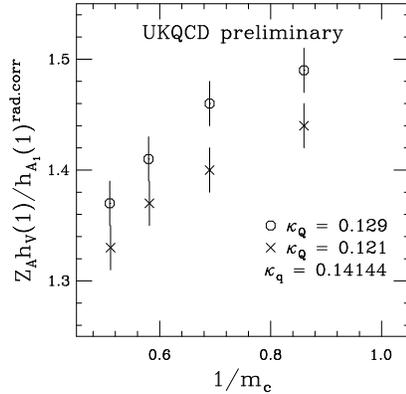

Figure 4. Ratio of $h_V/h_{A_1}$ normalised to $Z_A$.

## 4. Spin symmetry - ratios of form factors

With the exception of $h_+$ and $h_{A_1}$ the form factors are not protected against $O(\frac{1}{m_Q})$ corrections at zero recoil and the predictive power is very limited. For $h_V$ we find deviations from the exact symmetry limit $h_V(1) = 1$ of order 30-40%, even though the lattice vector current is normalised using the relation $\langle P,p|V_\mu|P,p\rangle = 2p_\mu$.

Following Neubert[12], we study the ratio $R = \frac{h_V}{h_{A_1}}$, which can also be extracted experimentally. This ratio depends on one subleading IW function which, ultimately, we would like to extract from a lattice calculation.

Here, we wish only to display the dependence of symmetry breaking effects for the ratio R at zero recoil. In Figure 4 we plot $Z_A R(1)$ for both b quark masses and all four c quark masses. We find large $1/m_Q$ corrections to the exact symmetry limit $R(1) = 1$ of the order 30-40% (note that $Z_A \sim 1.1$ [10]). We see an increase of this effect with decreasing b and c masses. This result is in good agreement with a sum rule prediction $R(1) = 1.33 \pm 0.08 [5]$.

Furthermore, we study the ratio $h_+/h_{A_1}$ of the two form factors protected by Luke's Theorem. We find this ratio to be close to one at zero recoil, as expected, with a slight decrease for bigger $v \cdot v'$. This is another indication that spin symmetry holds for masses around that of the charm quark.

## 5. Conclusions

We have shown that flavour symmetry holds for $h_{A_1}$ for heavy quark masses around charm and that $O(\frac{1}{m_Q})$ corrections to $h_{A_1}$ are small for $v \cdot v' < 1.3$.

The ratio $h_+/h_{A_1}$ is close to one for $v \cdot v' < 1.3$ providing evidence for spin symmetry.

Both $h_V$ and $\frac{h_V}{h_{A_1}}$ receive large $O(\frac{1}{m_Q})$ corrections of order 30 to 40 %.

## 6. Acknowledgments

This work was carried out on a Meiko i860 Computing Surface supported by SERC grant GR/G32779, Meiko Limited, the University of Edinburgh and by SERC grant GR/H01069.This work was carried out on a Meiko i860 Computing Surface supported by SERC grant GR/G32779, Meiko Limited, the University of Edinburgh and by SERC grant GR/H01069.